\documentclass{appolb}
\usepackage{graphicx}
\usepackage{sidecap}
\usepackage{tikz}
\usepackage{floatrow}
\usepackage{microtype}
\usepackage{hyperref}
\usepackage[utf8]{inputenc} 

\def\Xint#1{\mathchoice
   {\XXint\displaystyle\textstyle{#1}}%
   {\XXint\textstyle\scriptstyle{#1}}%
   {\XXint\scriptstyle\scriptscriptstyle{#1}}%
   {\XXint\scriptscriptstyle\scriptscriptstyle{#1}}%
   \!\int}
   
\def\XXint#1#2#3{{\setbox0=\hbox{$#1{#2#3}{\int}$}
     \vcenter{\hbox{$#2#3$}}\kern-.5\wd0}}

\def\dashint{\Xint-}

\begin{document}
\title{ Renormalization group and scattering-equivalent Hamiltonians on a coarse momentum grid
\thanks{Presented by Mar\'ia G\'omez-Rocha at Excited QCD 2020, February 2-8, \\
Krynica Zdr\'oj, Poland. 
\\
Work supported by the Spanish MINECO and
    European FEDER funds (FIS2017-85053-C2-1-P), Junta de
    Andaluc\'{\i}a (grant FQM-225) and Juan de la Cierva-Incorporacion programme, Grant Agreement No. IJCI-2017-31531.}%
}
\author{Mar\'ia G\'omez-Rocha, Enrique Ruiz Arriola
\address{
Departamento de F\'isica At\'omica, Molecular y Nuclear 
\\
and Instituto Carlos I de
F\'isica Te\'orica y Computacional, Universidad de Granada,
Avenida de Fuente Nueva s/n, 18071 Granada, Spain.
}
}
\maketitle
\begin{abstract}

We consider the $\pi\pi$-scattering problem in the context of the Kadyshevsky equation.
In this scheme, we introduce a momentum grid and provide an isospectral definition of the phase-shift based on the spectral shift of a Chebyshev angle. 
We address the problem of the unnatural high momentum tails present in the fitted interactions which reaches energies far beyond the maximal center-of-mass energy of $\sqrt{s}=1.4$ GeV. It turns out that these tails can be integrated out by using a block-diagonal generator of the SRG.

\end{abstract}
  
\section{Introduction}

Scattering phase shifts is one of the most important sources of information on the interaction between particles. In general, their theoretical calculation amounts to solving integral equations such as the Lippmann-Schwinger (LS) equation \cite{Lippmann:1950zz} in the non relativistic case or the Bethe-Salpeter equation~\cite{Salpeter:1951sz} (BSE) in the relativistic one. 
In any case, a potential describing the interaction is required and, in general, the integrals involved must be calculated numerically.  
The quality of the approximation implicit in the numerical method employed is of crucial relevance for ensuring the desired predictive power of the approach. 

Hamiltonian methods are very useful when treating hadronic reactions as a few-body problem by means of the corresponding potentials.  They have the advantage that the observables do not depend on the basis used as the spectrum is invariant under unitary transformations.
Separable potentials provide numerous advantages when solving the corresponding equations.
It often occurs, however, that separable potential that fit the interaction display an annoying long momentum tail, not natural to the physical problem. 

In this work we consider the $\pi\pi$-scattering problem and handle the long-momentum tail issue by means of the similarity renormalization group (SRG) for Hamiltonians~\cite{Glazek:1993rc,Glazek:1994qc}. We present, furthermore, a new alternative method for the calculation of phase shifts in a Hamiltonian framework~\cite{Gomez-Rocha:2019xum}.  

\section{Phase shifts and the spectral-shift method}

Among the 3d-reductions of the BSE, the Kadyshevsky equation~\cite{Kadyshevsky:1967rs}, is of particular interest in the context of $\pi\pi$ interactions, as it can be easily extended to the three-body problem. The reaction matrix in the Kadyshevsky scheme is given by
\begin{eqnarray}
R_l (p',p,\sqrt{s}) &=& V_l(p',p) 
\ + \ 
\dashint_0^\infty dq \, \frac{q^2}{4 E_q^2} 
\frac{V_l(p',q) R_l (q,p, \sqrt{s})}{\sqrt{s}-2 E_q }  \ ,
\label{eq:Kad-PV}
\end{eqnarray}
where $l$ is the orbital angular momentum, $s$ is the Mandelstam variable, and the subscript $q$ (or, analogously, $p$) in $E_q$ indicates that $E_q=\sqrt{\vec q^2 + m^2}$.
Scattering phase shifts $\delta_l(p)$ can be calculated from 
\begin{eqnarray}
  -\tan \delta_l(p) &=& \frac{\pi}{8} \frac{p}{E_p} R_l (p,p,\sqrt{s} ) \ .
\end{eqnarray}

The Hamiltonian version of  Eq.~(\ref{eq:Kad-PV})  is the eigenvalue equation
\begin{eqnarray}
  H \Psi_l(p)  \equiv 2 E_p \Psi_l(p) + \int_0^\infty dq \frac{q^2}{4 E_q^2} V_l(p,q)
  \Psi_l (q)  \ = \ E \, \Psi_l(p)  \ .
  \label{eq:Kad-Hamiltonian}
\end{eqnarray}

In general, it needs to be solved numerically. We choose the Gauss-Chebyshev quadrature: 
\begin{eqnarray}
p_n &=& \frac{\Lambda_{\rm num}}{2}\left[ 1- \cos \phi_n \right] \ , \quad 
w_n \ = \   \frac{\Lambda_{\rm num}}{2} \frac{\pi}{N} \phi_n \  ,
\quad \phi_n={\pi\over N}\left(n - {1\over 2}\right)
\label{eq:pnwn}
\end{eqnarray}
where $n=1, \dots, N$.  The variable $\phi_n$ is the 
{\it Chebyshev angle}, and $\Lambda_{\rm num}=p_{\rm{max}}$ is a high-momentum UV cutoff. 

We consider the separable model potential given in Ref.~\cite{Mathelitsch:1986ez}, which fits experimental data in the low-energy regime. One notices that such potentials present long-momentum tails that reach energies beyond the experimental region. This issue will be addressed in the next section.  

On the momentum grid, Eq.~(\ref{eq:Kad-Hamiltonian}) becomes 
\begin{eqnarray}
2 E_n \Psi_n 
  + \sum_k w_k \frac{p_k^2}{4 E_k^2} V_{n,k} \Psi_k = \sqrt{s} \Psi_n  \ .
\label{eq:KadH}  
\end{eqnarray}

It is already known that phase shifts can be determined from the spectrum of the Hamiltonian,
as it has been explained by DeWitt~\cite{DeWitt:1956be} and Fukuda and
Newton~\cite{Fukuda:1956zz}. They related the shift produced in the spectrum after introducing the interaction to the scattering phase
shifts appearing in the $S$-matrix. 
Such a relation is based on the use
of an equidistant energy and momentum grids, respectively. Since our Gauss-Chebyshev grid is equidistant in the Chebyshev angle, the
corresponding relation is~\cite{Gomez-Rocha:2019xum,Gomez-Rocha:2019rpj}
\begin{eqnarray}
  \delta_n= - \pi \frac{\Phi_n- \phi_n}{\Delta \phi_n} \equiv - \pi \frac{\Delta \Phi_n}{\Delta \phi_n} \ ,
\label{eq:phishift}  
\end{eqnarray}
where,
$\Delta \phi_n={\pi\over N}$, and the ``interacting'' angles $\Phi_n$ are
calculated form the Hamiltonian eigenvalue
$\sqrt{s}=2E_n=\sqrt{m^2+P_n^2}$, inverting  Eqs.~(\ref{eq:pnwn}) after replacing $p_n$ by $P_n$.

\begin{figure*}[h]

\includegraphics[scale=0.45]{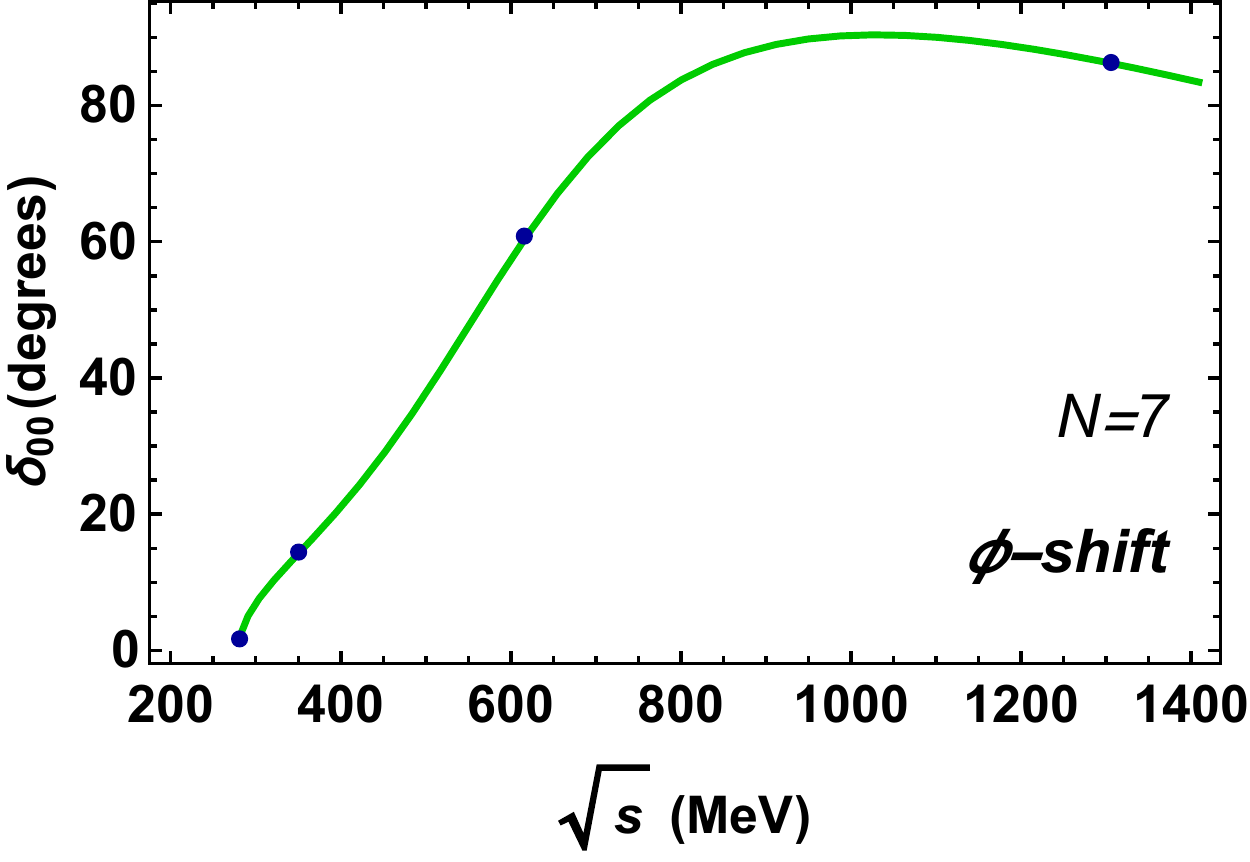}
\includegraphics[scale=0.45]{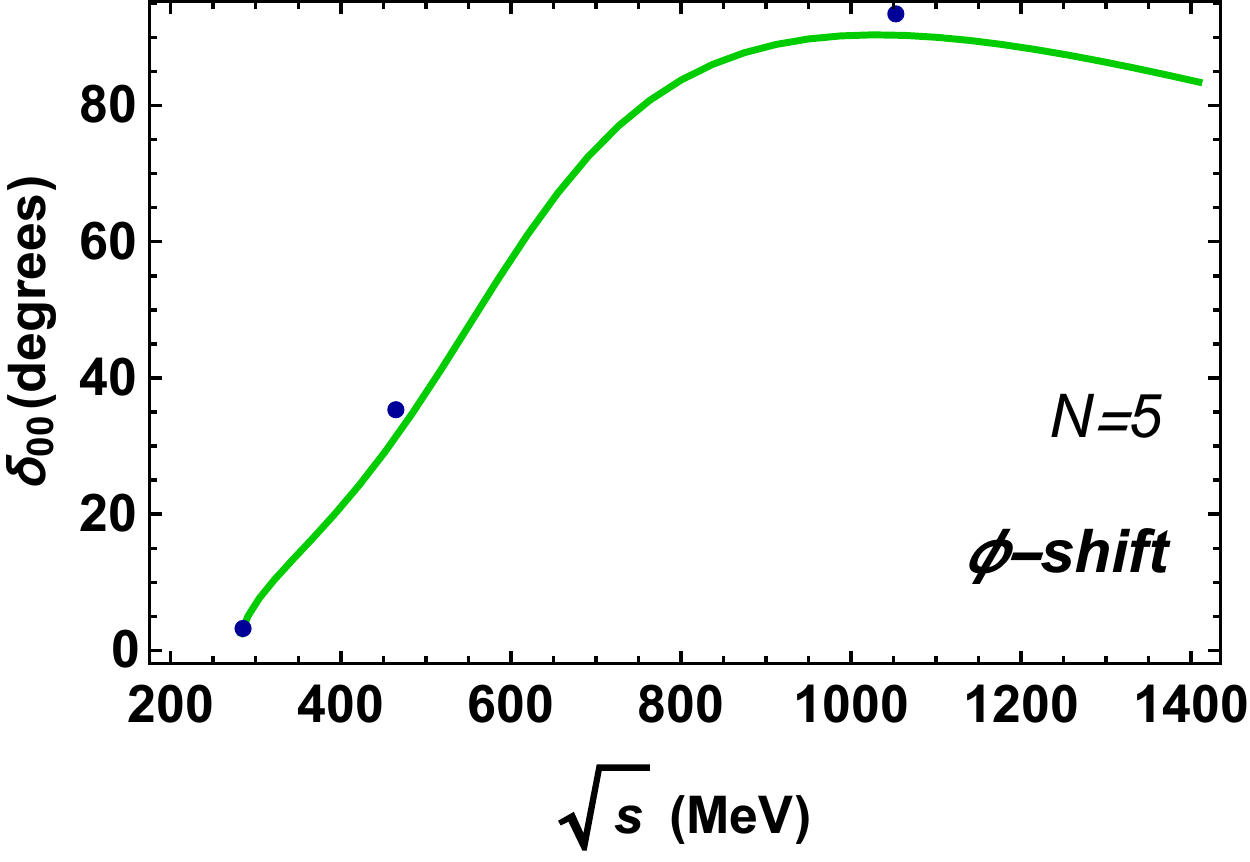}

\floatbox[{\capbeside\thisfloatsetup{capbesideposition={right,center},capbesidewidth=5.5cm}}]{figure}[\FBwidth]
{\caption{$\pi\pi$ scattering phase shifts in the $S_0$ channel calculated with the $\phi$-shift method using a grid of $N=$7, 5, and 4 points (blue dots) compared with the model fit (green line).}\label{fig:pocospuntos}}
{
\includegraphics[scale=0.45]{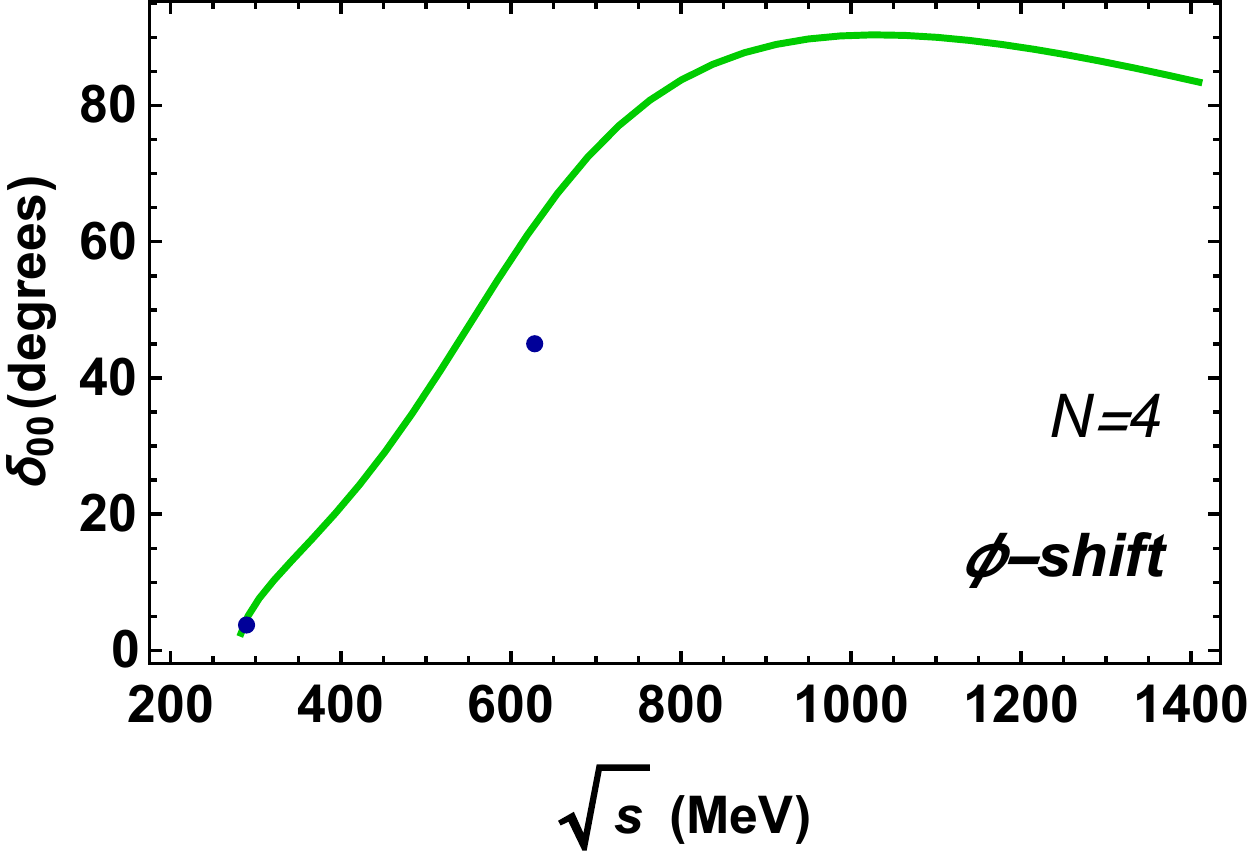}}
\end{figure*}

Figure~\ref{fig:pocospuntos} shows scattering phase shifts in the $LI=$00 channel calculated using Eq.~(\ref{eq:phishift}) for different number of grid points,  compared with the exact solution that fits the data in the inelastic range\footnote{Cf. also the case with a larger number of points in the left panel in figure~\ref{fig:longrange}.}. The calculation starts slightly differing form the exact solution for a grid of 5 points. This is an extremely small number as it can be noticed by comparing with results obtained using other standard methods, such as e.g. the Lippmann-Schwinger equation~\cite{Gomez-Rocha:2019rpj}.

\section{Similarity renormalization group and scattering}

\begin{figure}[h]
    \centering
\includegraphics[scale=0.45]{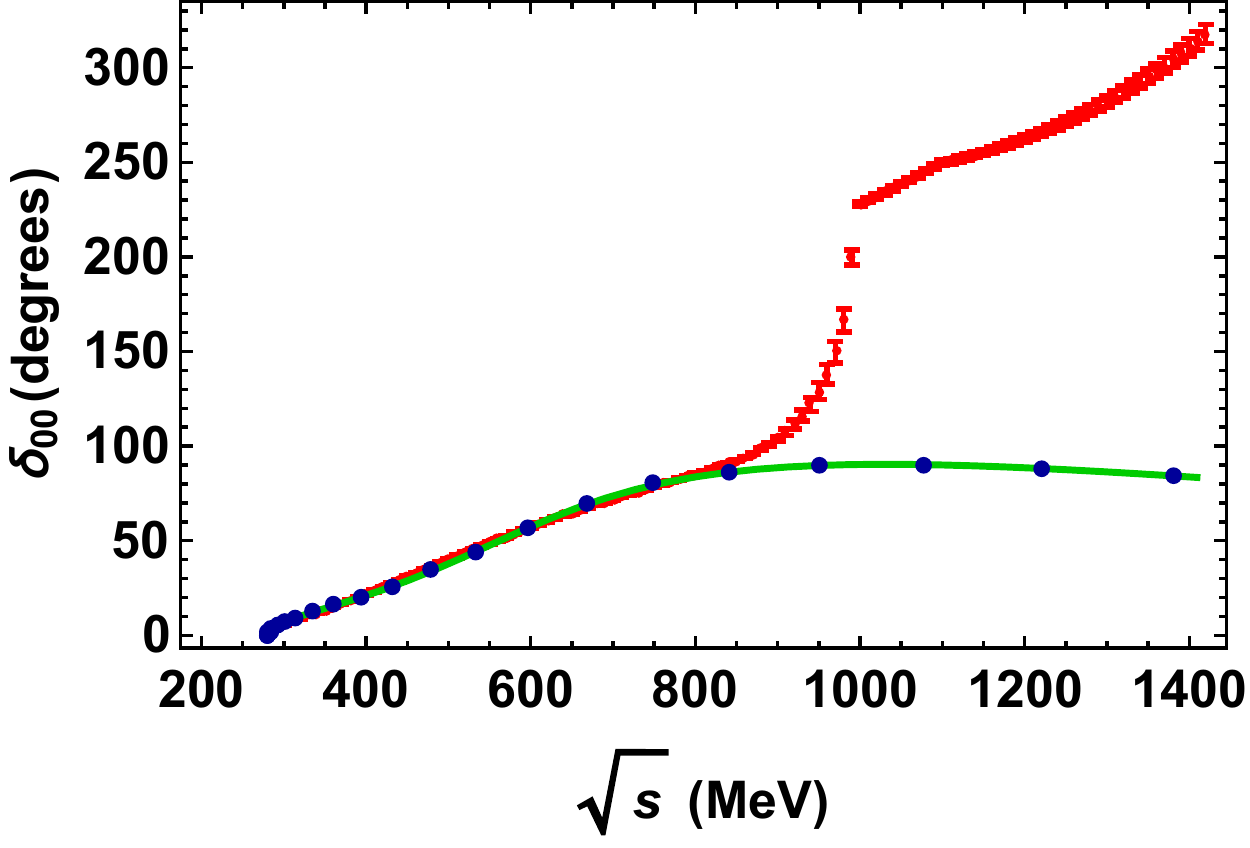}
\includegraphics[scale=0.47]{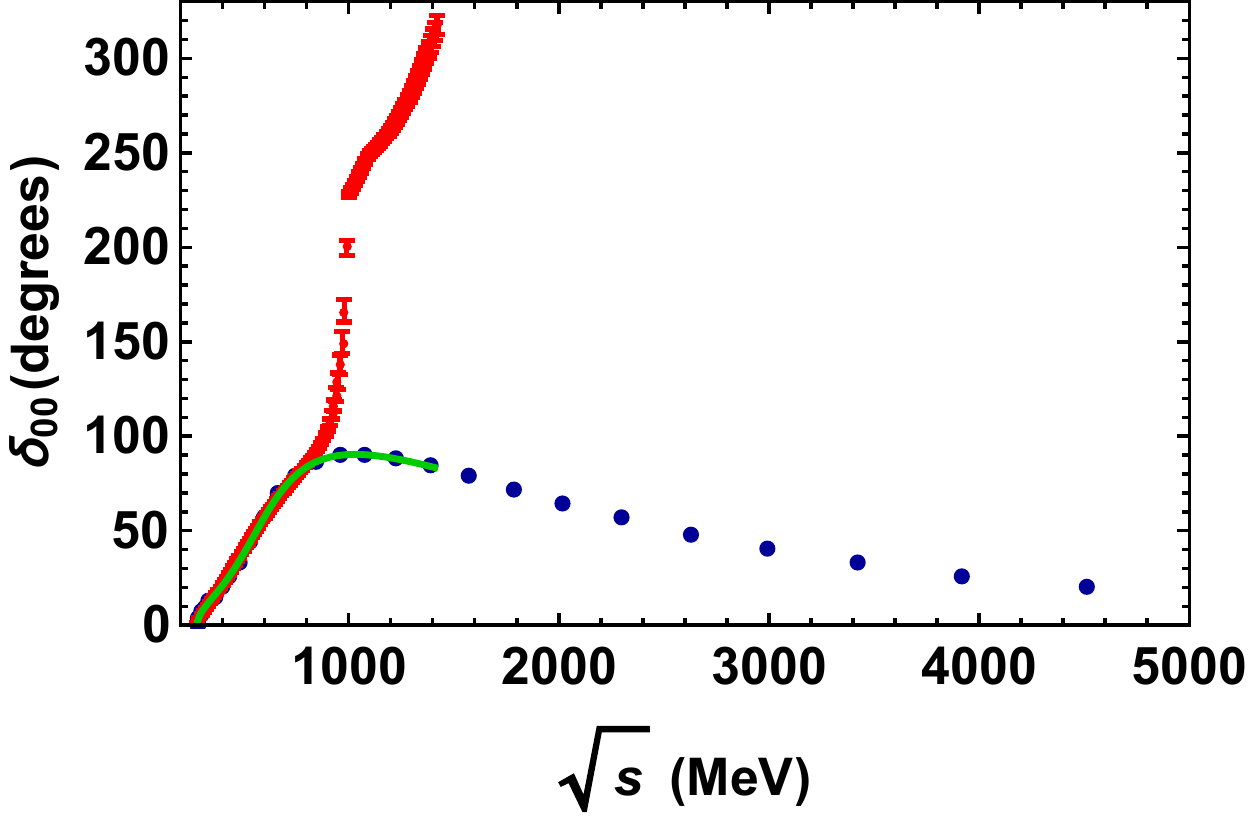}
    \caption{Phase shifts calculated using the $\phi$-shift method (blue dots) compared with the experimental data~\cite{GarciaMartin:2011cn} (red dots with error bars) and with the model fit (solid, green line). The left panel shows the experimental region. The right panel shows the energy region covered by the Hamiltonian.}
    \label{fig:longrange}
\end{figure}

In order to calculate the phase shifts in the experimental range shown in Figure~\ref{fig:pocospuntos}, we used the separable model potentials given in Ref.~\cite{Mathelitsch:1986ez}, which fit the experimental data in the range of interest. The annoying fact is that such potentials present very long tails (even up to 30 GeV), which involve energies far beyond the physical problem (only $\sim$1.5 GeV). In fact, in order to describe scattering phase shifts in the range given in the left panel of figure~\ref{fig:longrange}, one needs to calculate them in the entire range reached by the potential, as it is illustrated in the right panel of the same figure. 

This issue is not only unnatural from the physical point of view, it implies also an undesired computational load which does not provide any additional information about the interaction.  

We address this issue by means of the SRG~\cite{Glazek:1993rc,Glazek:1994qc,Wilson:1994fk}. This procedure is based on the idea that it is possible to apply a (scale dependent) unitary transformation on a given initial Hamiltonian, in such a way that after a continuous evolution along the running of the scale parameter, the Hamiltonian is transformed into a more convenient basis. Similarity transformations preserve the eigenvalues and, as a consequence, scattering phase shifts obtained from the spectrum remain invariant. 
The SRG evolution of the Hamiltonian obeys 
\vspace{-0.5cm}
\begin{eqnarray}
\frac{d H_t}{dt}= [[G_t,H_t],H_t] \ , 
\label{eq:srg}
\end{eqnarray}
where $t$ is the renormalization-group parameter and $G_t$ is the generator which determines the basis into which the {\it running} Hamiltonian, $H_t$, is transformed\footnote{Although the SRG has been mostly used in nuclear physics, recent progress has been made in the last few years concerning its application to QCD~\cite{Gomez-Rocha:2015esa,Glazek:2017rwe,Serafin:2018aih} .}.
Different choices of $G_t$ involve different bases for the effective Hamiltonian. In this work, we consider a bloc-diagonal generator, with the structure  $G=PHP+QHQ$, where $P$ and $Q$ are the orthogonal projectors, $P=\theta(\Lambda-p)$ and $Q=
\theta(p-\Lambda)$. This transformation defines two subspaces separated by a cutoff $\Lambda$, and converts the initial Hamiltonian into a block-diagonal matrix. The cutoff ($\Lambda\sim$1400 GeV) is chosen in such a way that the energy range beyond the experimental region decouples form the other one. The matrix elements relevant to the physical problem remain now in the small block and the scattering phase shifts can be calculated from the eigenvalues of this reduced matrix. 

\begin{figure*}[h]

\includegraphics[scale=0.45]{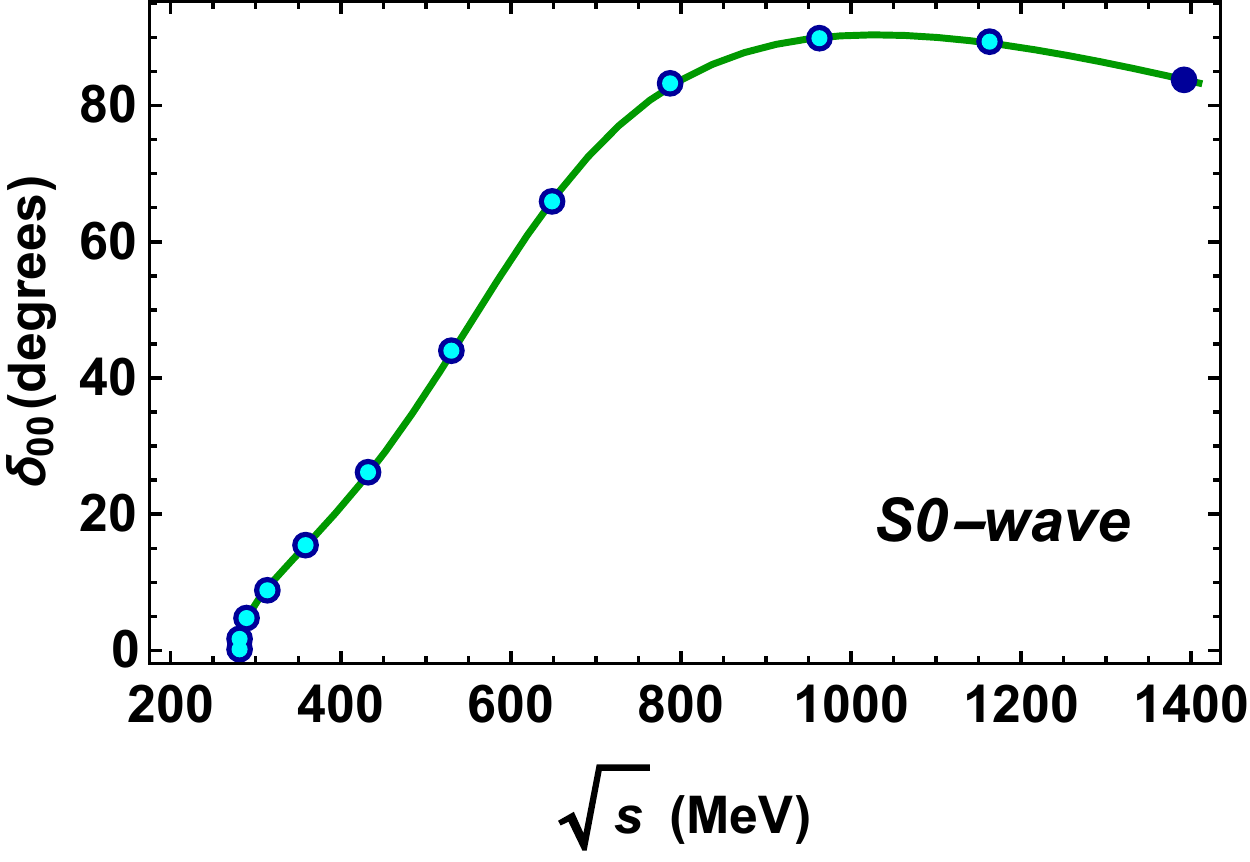}
\includegraphics[scale=0.45]{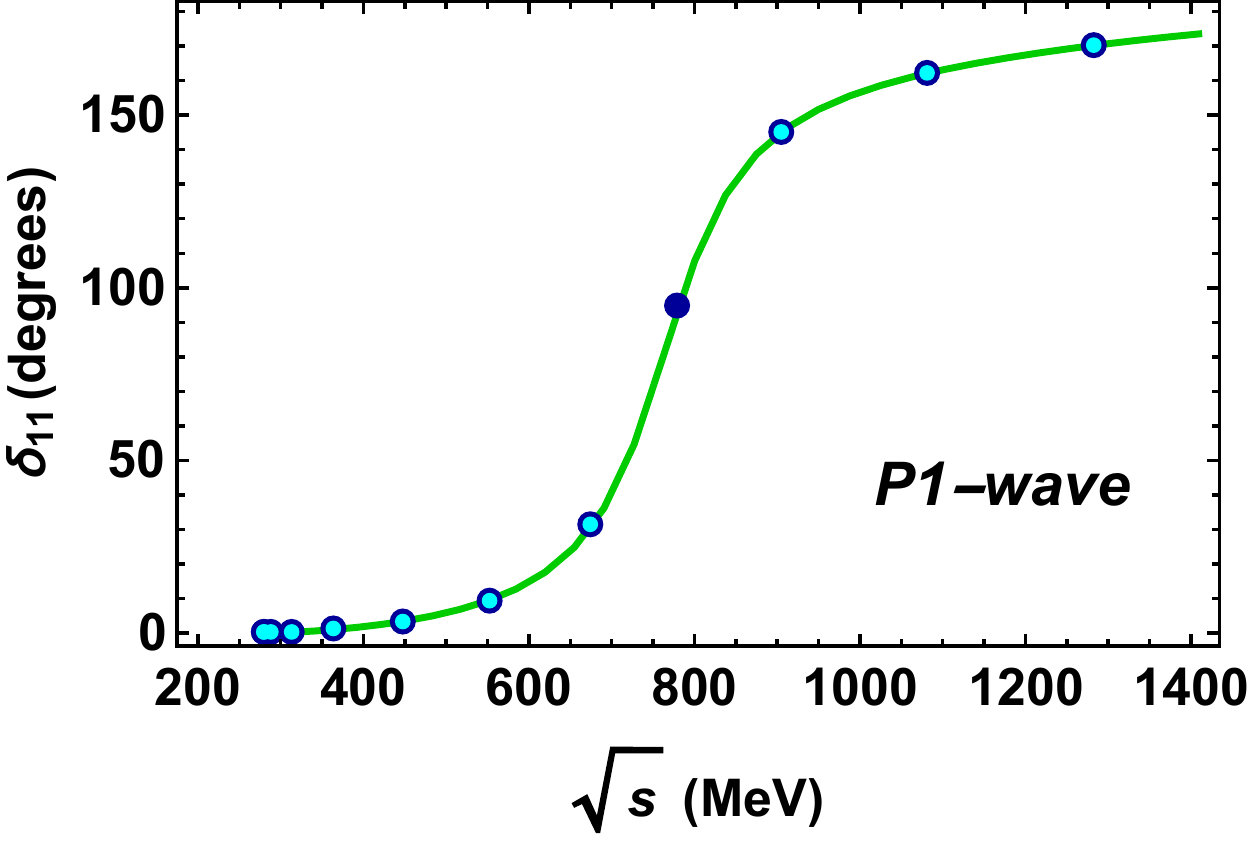}

\floatbox[{\capbeside\thisfloatsetup{capbesideposition={left,center},capbesidewidth=5.5cm}}]{figure}[\FBwidth]
{\caption{Phase shifts in different $\delta_{lI}$ channels calculated with the $\phi$-shift method using the original Hamiltonian (dark blue dots) and with the block below the cutoff (light blue dots), compared with the model fit (solid, green line). The calculation has been made with an initial grid of 50 points.}\label{fig:equivalent}}
{
\includegraphics[scale=0.45]{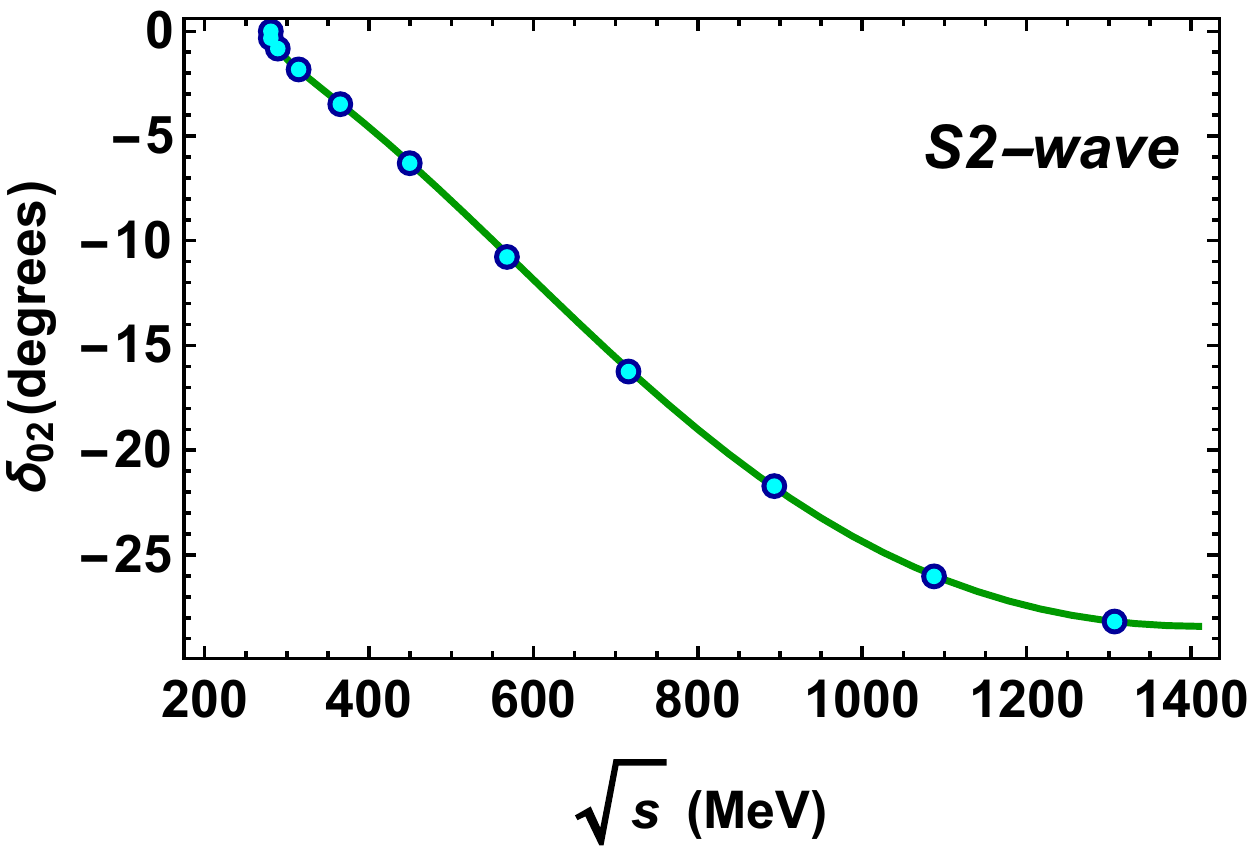}}
\end{figure*}

Figure~\ref{fig:equivalent} shows the phase shifts calculated in channels $lI=$00, 11, and 02, using the $\phi$-shift method with the initial $50\times 50$  matrix (dark, larger dots) and with the small block ($11\times 11$ matrix) of the evolved Hamiltonian (lighter blue dots). While the results are the same, the number of computational operations needed to obtain them is much smaller in the latter case.

\vspace{-0.5cm}

\section{Conclusions}

In this work we have presented two different methods that we have combined in a convenient way in order to calculate scattering phase shifts in a remarkable simple fashion and with a minimal computational cost.

On the one hand, we have presented the newest version of the so-called spectral-shift methods for the calculation of scattering phase shifts.
Alike the energy-shift~\cite{DeWitt:1956be} and the momentum-shift method~\cite{Fukuda:1956zz}, it is based on the displacement of the grid associated with the integration variable, that occurs after the interaction, in an analogous way to what happens in the energy levels appearing in the scattering process in a box. 

The $\phi$-shift method~\cite{Gomez-Rocha:2019xum}, suitable to the Chebyshev momentum grid employed here, yields results with exceptional accuracy, even when one considers a grid with a small number o points. The method appears as a useful tool, that is computationally cheaper than other conventional methods. 

On the other hand, we have used the SRG method to transform the original Hamiltonian matrix into a similar one (with the same eigenvalues) with block-diagonal structure. This effective Hamiltonian distinguishes two energy regions separated by a cutoff. This allows us to select the experimental region and proceed with the calculation of scattering phase shifts using only the matrix elements of the block below the cutoff. This procedure simplifies even more the calculation.

To construct the Hamiltonian employed in all these calculations, we have chosen the Kadyshevsky scheme, in which the three-body problem is easy to implement. 
Once the $\pi\pi$ interaction is understood in this framework, one can start considering the study of $\pi\pi\pi$ interaction appearing, i.e. in the $\omega$
or the $A_1$ decays, among other problems.

\bibliography{2refs,newrefs}

\begin{thebibliography}{10}

\bibitem{Lippmann:1950zz}
B.A. Lippmann and J. Schwinger,
\newblock Phys. Rev. 79 (1950) 469.

\bibitem{Salpeter:1951sz}
E.E. Salpeter and H.A. Bethe,
\newblock Phys. Rev. 84 (1951) 1232.

\bibitem{Glazek:1993rc}
S.D. Glazek and K.G. Wilson,
\newblock Phys. Rev. D48 (1993) 5863.

\bibitem{Glazek:1994qc}
S.D. Glazek and K.G. Wilson,
\newblock Phys. Rev. D49 (1994) 4214.

\bibitem{Gomez-Rocha:2019xum}
M. Gómez-Rocha and E. Ruiz~Arriola,
\newblock Phys. Lett. B800 (2020) 135107.

\bibitem{Kadyshevsky:1967rs}
V.G. Kadyshevsky,
\newblock Nucl. Phys. B6 (1968) 125.

\bibitem{Mathelitsch:1986ez}
L. Mathelitsch and H. Garcilazo,
\newblock Phys. Rev. C 32 (1985) 1635.

\bibitem{DeWitt:1956be}
B.S. DeWitt,
\newblock Phys. Rev. 103 (1956) 1565.

\bibitem{Fukuda:1956zz}
N. Fukuda and R.G. Newton,
\newblock Phys. Rev. 103 (1956) 1558.

\bibitem{Gomez-Rocha:2019rpj}
M. Gómez-Rocha and E. Ruiz~Arriola,
\newblock Phys. Rev. D 101 (2020) 036003, 1911.08990.

\bibitem{GarciaMartin:2011cn}
R. Garcia-Martin et~al.,
\newblock Phys. Rev. D83 (2011) 074004, 1102.2183.

\bibitem{Wilson:1994fk}
K.G. Wilson et~al.,
\newblock Phys. Rev. D49 (1994) 6720, hep-th/9401153.

\bibitem{Gomez-Rocha:2015esa}
M. Gómez-Rocha and S.D. Głazek,
\newblock Phys. Rev. D92 (2015) 065005, 1505.06688.

\bibitem{Glazek:2017rwe}
S.D. Głazek et~al.,
\newblock Phys. Lett. B773 (2017) 172, 1705.07629.

\bibitem{Serafin:2018aih}
K. Serafin et~al.,
\newblock Eur. Phys. J. C 78 (2018) 964, 1805.03436.

\end{thebibliography}

\bibliographystyle{h-elsevier}

\end{document}